# Quantifying the nonadiabaticity strength constant in recently discovered highly-compressed superconductors

Evgeny F. Talantsev [1,2]

[1]M. N. Miheev Institute of Metal Physics, Ural Branch, Russian Academy of Sciences, 18, S. Kovalevskoy St., Ekaterinburg 620108, Russia
[2]NANOTECH Centre, Ural Federal University, 19 Mira St., Ekaterinburg 620002, Russia

## Abstract

Superconductivity in highly-pressurized hydrides became primary direction for the exploration of fundamental upper limit for the superconducting transition temperature, $T_c$, after Drozdov *et al* (*Nature* **2015**, *525*, 73) discovered superconducting state with $T_c = 203 \, K$ in highly-compressed sulphur hydride. To date several dozens of high-temperature superconducting polyhydrides have been discovered. In addition, recently, it was reported that highly-compressed titanium and scandium exhibit record-high $T_c$ (up to 36 K), which is by manifold exceeded $T_c = 9.2 \, K$ of niobium, which is the record high-$T_c$ ambient pressure metallic superconductor. Here we analysed experimental data on for recently discovered high-pressure superconductors (which exhibit high transition temperatures within their classes): elemental titanium (Zhang *et al*, *Nature Communications* **2022**; Liu *et al*, *Phys. Rev. B* **2022**), $TaH_3$ (He *et al*, *Chinese Phys. Lett.* **2023**), $LaBeH_8$ (Song *et al*, *Phys. Rev. Lett.* **2023**), and black (Li *et al*, *Proc. Natl. Acad. Sci.* **2018**) and violet (Wu *et al*, *arXiv* **2023**) phosphorous, to reveal the nonadiabaticity strength constant, $\frac{T_\theta}{T_F}$ (where $T_\theta$ is the Debye temperature, and $T_F$ the Fermi temperature) in these superconductors. The analysis showed that $\delta$-phase of titanium and black phosphorous exhibit the $\frac{T_\theta}{T_F}$ which are nearly identical to ones associated in A15 superconductors, while studied hydrides and violet phosphorous exhibit the constants in the same ballpark with $H_3S$ and $LaH_{10}$.

**Quantifying the nonadiabaticity strength constant in recently discovered highly-compressed superconductors**

## 1. Introduction

The discovery of near-room temperature superconductivity in highly compressed sulphur hydride by Drozdov *et al* [1] manifested a new era in superconductivity. This research field represents one of the most fascinating scientific and technological exploration in modern condensed matter physics where advanced first principles calculations [2–11] are essential part of the experimental quest for the discovery of new hydrides phases [12–21], and both of these directions drive the development of new experimental techniques to study highly-pressurized materials [22–31].

From 2015 till now, several dozens of high-temperature superconducting polyhydride phases have been discovered and studied [1], [12–21], [24,32–40], [41–43]. At the same time, high-pressure studies of the superconductivity in non-hydrides are also progressed recently [44–53], including observation of $T_c > 26\ K$ in highly-compressed elemental titanium [54,55] and scandium [56,57], and $T_c^{onset} \cong 78\ K$ in $La_3Ni_2O_7$ [58].

First principles calculations [12–21],[59–68] are essential tool in the quest for room-temperature superconductivity (which was used [65] to explain experimental result [69] for one of the most difficult to explain hydride case, $AlH_3$), and primary calculated parameter in these calculations is the transition temperature, $T_c$. As, the confirmation of the predicted $T_c$, as the determination of other fundamental ground state parameters, for instance, the upper critical field, $B_{c2}(0)$ [5,24,33,39], the lower critical field, $B_{c1}(0)$ [12,22], the self-field critical current density, $J_c(sf,T)$ [24,70–72], the London penetration depth, $\lambda(0)$ [22,23,73,74], the superconducting energy gap amplitude, $\Delta(0)$ [75–77], and gap symmetry [78,79], etc., are the task for experiment and data analysis

Another complication in understanding of the superconductivity in highly-pressurized materials is the phenomenon of nonadiabaticity, which originates from a fact that Migdal-Eliashberg theory of the electron-phonon mediated superconductivity [80,81] is based on primary assumption/postulate that the superconductor obeys the inequality:

$$T_\theta \ll T_F \qquad (1)$$

where, $T_\theta$ is the Debye temperature, and $T_F$ is the Fermi temperature. In other words, Eq. 1 implies that the superconductor exhibits fast electric charge carriers and slow ions. This assumption simplifies theoretical model of the electron-phonon mediated superconductivity, however, Eq. 1 is not satisfied for many unconventional superconductors [82–90] (which was first pointed out by Pietronero and co-workers [91–94]) and many highly-compressed superconductors [79,89,95–97].

While theoretical aspects of the non-adiabatic effects can be found elsewhere [11,82,88,91–95], in practice, the strength of the nonadiabatic effects can be quantified by the $\frac{T_\theta}{T_F}$ ratio [89,90] for which in Ref. [89] three characteristic ranges were proposed:

$$\begin{cases} \frac{T_\theta}{T_F} < 0.025 \rightarrow adiabatic\ superconductor; \\ 0.025 \lesssim \frac{T_\theta}{T_F} \lesssim 0.4 \rightarrow moderately\ strong\ nonadiabatic\ superconductor; \\ 0.4 < \frac{T_\theta}{T_F} \rightarrow nonadiabatic\ superconductor. \end{cases} \qquad (2)$$

It was found in Ref. [89], and confirmed in Ref. [79], that superconductors with $T_c > 10\ K$ (from a dataset of 46 superconductors from all major superconductors families) exhibit the $\frac{T_\theta}{T_F}$ ratio in the range $0.025 \lesssim \frac{T_\theta}{T_F} \lesssim 0.4$. This is interesting and theoretically unexplained empirical observation.

In this study we further extended empirical $\frac{T_\theta}{T_F}$ database by deriving several fundamental parameters:

(1) the Debye temperature, $T_\theta$;

(2) the electron-phonon coupling constant, $\lambda_{e-ph}$;

(3) the ground state coherence length, $\xi(0)$;

(4) the Fermi temperature $T_F$;

(5) the nonadiabaticity strength constant, $\frac{T_\theta}{T_F}$;

(6) and the ratio $\frac{T_c}{T_F}$;

for five recently discovered highly-compressed superconductors for which reported raw experimental data are enough to deduce mentioned above parameters, and which represent materials with high or record high $T_c$ in their families:

(1) elemental titanium, $\delta - Ti$ [54,55];

(2) $TaH_3$ [21];

(3) $LaBeH_8$ [98];

(4) black phosphorous [99–101];

(5) violet phosphorous [53].

In the result, we derived the nonadiabaticity strength constant, $\frac{T_\theta}{T_F}$, for these superconductors and confirmed previously reported empirical observation [79,89] that superconductors with $T_c > 10\ K$ obey the condition $0.025 \lesssim \frac{T_\theta}{T_F} \lesssim 0.4$.

## 2. Utilized models and data analysis tools

### 2.1. Debye temperature

Debye temperature, $T_\theta$, is one of fundamental parameters which determines the superconducting transition temperature, $T_c$, within electron-phonon phenomenology [81,102–106]. This parameter can be deduced as a free-fitting parameter from a fit of temperature dependent resistance, $R(T)$, to the saturated resistance model within the Bloch-Grüneisen (BG) equation [107–110]:

$$R(T) = \cfrac{1}{\cfrac{1}{R_{sat}} + \cfrac{1}{R_0 + A\left(\frac{T}{T_\theta}\right)^5 \int_0^{\frac{T_\theta}{T}} \frac{x^5}{(e^x - 1)(1 - e^{-x})} dx}} \tag{3}$$

where $R_{sat}$, $R_0$, $T_\theta$ and $A$ are free fitting parameters.

## 2.2. The electron-phonon coupling constant

From the deduced $T_\theta$ and measured $T_c$, which we defined by as strict as possible resistance criterion of $\frac{R(T)}{R_{norm}} \to 0$, where $R_{norm}$ is the sample resistance at the onset of the superconducting transition, the electron-phonon coupling constant, $\lambda_{e-ph}$, can be calculated as the root of advanced McMillan equation [103–106]:

$$T_c = \left(\frac{1}{1.45}\right) \times T_\theta \times e^{-\left(\frac{1.04\left(1 + \lambda_{e-ph}\right)}{\lambda_{e-ph} - \mu^*\left(1 + 0.62\lambda_{e-ph}\right)}\right)} \times f_1 \times f_2^* \tag{4}$$

$$f_1 = \left(1 + \left(\frac{\lambda_{e-ph}}{2.46(1 + 3.8\mu^*)}\right)^{3/2}\right)^{1/3} \tag{5}$$

$$f_2^* = 1 + (0.0241 - 0.0735 \times \mu^*) \times \lambda_{e-ph}^2 \tag{6}$$

where $\mu^*$ is the Coulomb pseudopotential parameter, which we assumed to be $\mu^* = 0.13$ (which is typical value utilized in the first principles calculation for many electron-phonon mediated superconductors [54,111]).

## 2.3. Ground state coherence length

To deduce the ground state coherence length, $\xi(0)$, we fitted the upper critical field datatset, $B_{c2}(T)$, to analytical approximant of the Werthamer-Helfand-Hohenberg model [112,113], which was proposed by Baumgartner *et al* [114]:

$$B_{c2}(T) = \frac{1}{0.693} \times \frac{\phi_0}{2\pi \xi^2(0)} \times \left(\left(1 - \frac{T}{T_c}\right) - 0.153 \times \left(1 - \frac{T}{T_c}\right)^2 - 0.152 \times \left(1 - \frac{T}{T_c}\right)^4\right) \tag{7}$$

where $\phi_0 = \frac{h}{2e}$ is the superconducting flux quantum, $h = 6.626 \times 10^{-34} \, J \cdot s$ is Planck constant, $e = 1.602 \times 10^{-19} \, C$, and $\xi(0)$ and $T_c \equiv T_c(B=0)$ are free fitting parameters.

### 2.4. The Fermi temperature

Simplistic approach to calculate the Fermi temperature, $T_F$, is to use the expression of free-electron model [115,116]:

$$T_F = \frac{\varepsilon_F}{k_B} = \frac{\left(3\pi^2 n \hbar^3\right)^{\frac{3}{2}}}{2m_e(1+\lambda_{e-ph})k_B} \tag{8}$$

where $m_e = 9.109 \times 10^{-31} \, kg$ is bare electron mass, $\hbar = 1.055 \times 10^{-34} \, J \cdot s$ is reduced Planck constant, $k_B = 1.381 \times 10^{-23} \, m^2 \cdot kg \cdot s^{-2} \cdot K^{-1}$ is Boltzmann constant, and $n$ is the charge carrier density per volume ($m^{-3}$). Equation 8 can be used, if the Hall resistance measurements were analysed to estimate the charge carrier density, $n$.

If Hall resistance measurements were not performed, then to calculate the Fermi temperature, we utilized the equation [58,59]:

$$T_F = \frac{\pi^2 m_e}{8k_B} \times \left(1+\lambda_{e-ph}\right) \times \xi^2(0) \times \left(\frac{\alpha k_B T_c}{\hbar}\right)^2 \tag{9}$$

where $\alpha = \frac{2\Delta(0)}{k_B \cdot T_c}$ is the gap-to-transition temperature ratio and this is the only unknown parameter in Eq. 6.

### 2.5. The gap-to-transition temperature ratio

To calculate the Fermi temperature by Eq. 9 there is a need to know $\alpha = \frac{2\Delta(0)}{k_B \cdot T_c}$. In this study, to determine $\alpha = \frac{2\Delta(0)}{k_B \cdot T_c}$ we utilized the following approach. Carbotte [111] collected various parameters for 32 electron-phonon mediated superconductors, which exhibit $0.43 \leq \lambda_{e-ph} \leq 3.0$ and $3.53 \leq \frac{2\Delta(0)}{k_B \cdot T_c} \leq 5.19$. In Figure 1 we presented the dataset reported by Carbotte in his Table IV [111]. The dependence $\frac{2\Delta(0)}{k_B \cdot T_c}$ vs $\lambda_{e-ph}$ can be approximate by linear function (Fig. 1) [117]:

$$\frac{2\Delta(0)}{k_B T_c} = C + D \times \lambda_{e-ph} \qquad (10)$$

where $C = 3.26 \pm 0.06$, and $D = 0.74 \pm 0.04$.

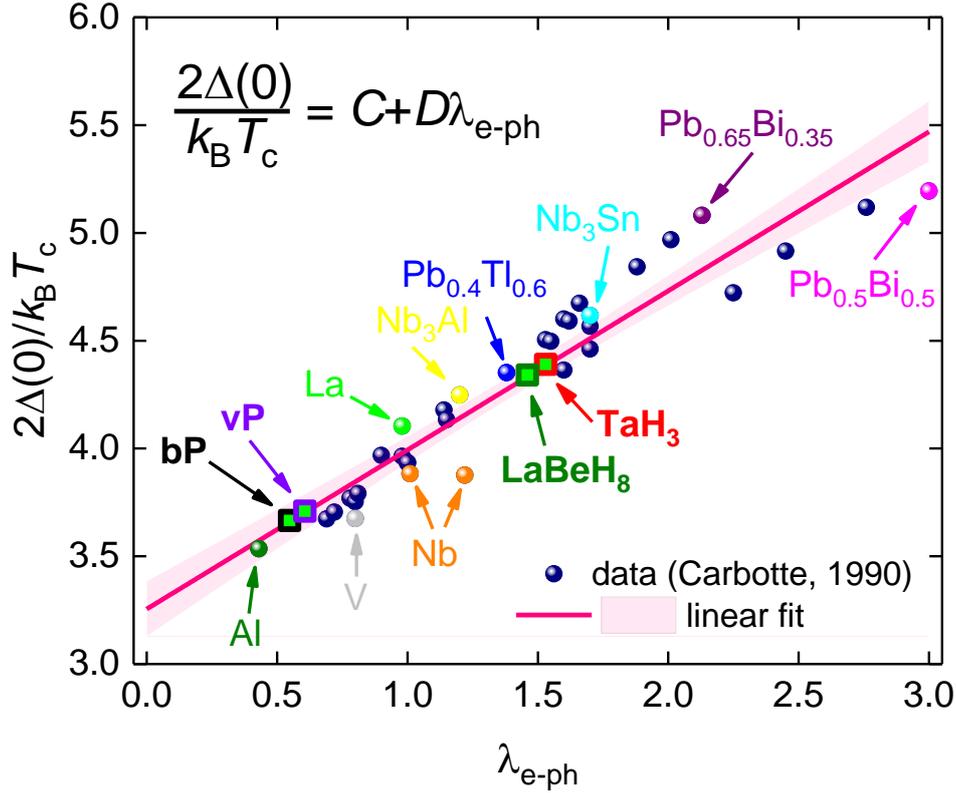

**Figure 1.** The gap-to-transition temperature ratio, $\frac{2\cdot\Delta(0)}{k_B \cdot T_c}$, vs the electron-phonon coupling constant, $\lambda_{e-ph}$, dataset reported by Carbotte in the Table IV of Ref. [111]. Linear fit is shown by pink line. Positions for some representative superconductors and superconductors studied in this report (where **bP** stands for black phosphorus and **vP** stands for violet phosphorus) are shown. 95% confidence bands for the linear fit are shown by pink shadow area.

As far as one can determine $\lambda_{e-ph}$ by utilized Equations (3)-(6), the $\frac{2\Delta(0)}{k_B \cdot T_c}$ ratio can be estimated from the Eq. (10).

## 3. Results

### 3.1. Highly-compressed titanium

Zhang *et al* [54] and Liu *et al* [55] reported on record high $T_c$ in $\delta - Ti$ phase compressed at megabar pressures. In Figure 2 we showed the fit of the $R(T)$ dataset measured by Zhang *et al* [54] for the $\omega - Ti$ phase compressed at $P = 18\ GPa$ to Equation 3.

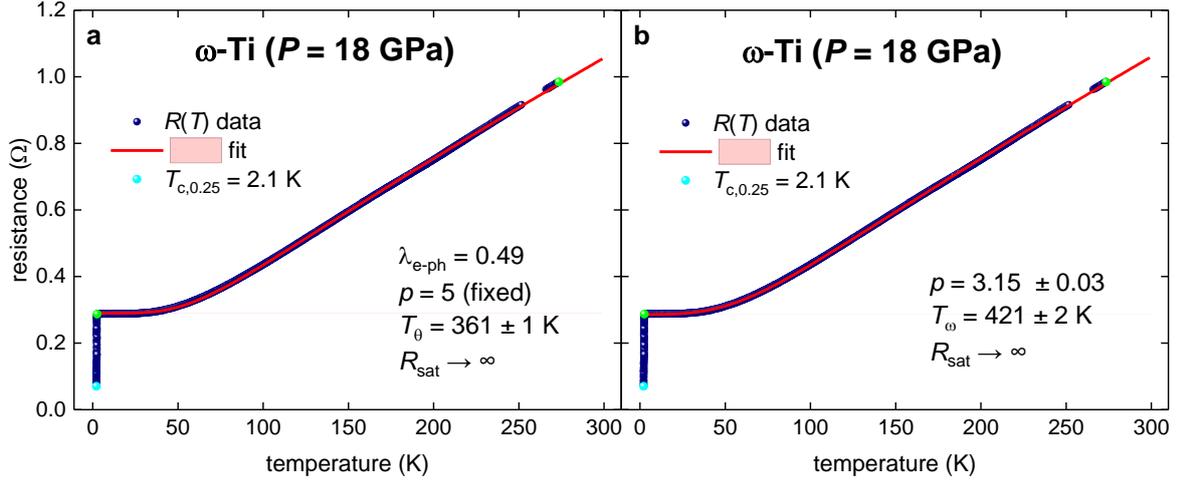

**Figure 2.** Temperature dependent resistance data, $R(T)$, for compressed titanium ($\omega - Ti$-phase at $P = 18$ GPa) and data fit to Eq. 3 (raw data reported by Zhang *et al* [54]). Green balls indicate the bounds for which $R(T)$ data was used for the fit to Eq. 3. (a) Fit to Debye model: $p = 5\ (fixed)$, $T_\theta = 361 \pm 1\ K$, $T_{c,0.25} = 2.1\ K$, $\lambda_{e-ph} = 0.49$, fit quality is 0.99988. (b) Fit to Eq. 3: $p = 3.15 \pm 0.03$, $T_\omega = 421 \pm 2\ K$, $T_{c,0.25} = 2.1\ K$, fit quality is 0.99995. 95% confidence bands are shown.

Deduced Debye temperature (Figure 2,a) for $\omega - Ti$-phase ($P = 18\ GPa$) is $T_\theta = 361 \pm 1\ K$ which is in ballpark value with $T_\theta(298\ K) = 380\ K$ for uncompressed pure elemental titanium, which exhibits $\alpha - Ti$ phase [118].

To calculate the electron-phonon coupling strength constant, $\lambda_{e-ph}$, by Equations 4-6, we defined the superconducting transition temperature, $T_c = 2.1\ K$, by the use of $\frac{R(T)}{R_{norm}} = 0.25$ criterion, which was chosen based on the lowest temperature, at which experimental $R(T)$ data measured at $P = 18\ GPa$ was reported by Zhang *et al* [54]. Deduced $\lambda_{e-ph} = 0.49$, which is very close to the $\lambda_{e-ph} = 0.43$ of pure elemental aluminium (Fig. 1, and Ref. [111]).

We also confirmed the power-law exponent $n = 3.1$ (reported by Zhang *et al* [54]) for the temperature dependent $R(T)$, which was extracted by Zhang *et al* [54] from the simple power-law fit of $R(T)$ at temperature range of $3\ K \leq T \leq 70\ K$:

$$R(T) = R_0 + C \times T^n \qquad (11)$$

where $R_0$, $C$ and $n$ are free fitting parameters. As we showed earlier [119], Eq. 10 does not always return correct $n$-values, and $R(T)$ data fit to Eq. 3, where $p$ is free-fitting parameter, is the reliable approach to derive the power-law exponent. However, for the given case, our fit to Eq. 3 (Figure, 2,b) returns the same power-law exponent, $p = 3.15 \pm 0.03$, to the one reported by Zhang *et al* [54].

In Figure 3 we showed $R(T)$ data measured by Zhang *et al* [54] and Liu *et al* [55] and data fits to Equations 3-7 for the $\delta - Ti$ phase compressed at $P = 154\ GPa$ (Figure 3,a), $P = 180\ GPa$ (Figure 3,b), $P = 183\ GPa$ (Figure 3,c), and $P = 245\ GPa$ (Figure 3,d).

While Liu *et al* [55] reported first principles calculation result for $\lambda_{e-ph}$ and logarithmic frequency $\omega_{log}$ for highly compressed titanium over wide range of applied pressure, in Figure 4 we presented a comparison of the deduced $\lambda_{e-ph}$ and $T_\theta$ values from experiment and calculated ones [55]. To compare $\omega_{log}$ (calculated by first principles calculations) and $T_\theta$ deduced from experiment, we used theoretical expression proposed by Semenok [120]:

$$\frac{1}{0.827} \times \frac{\hbar}{k_B} \times \omega_{log} \cong T_\theta \qquad (12)$$

In Figure 4,c we also show $T_F$ values calculated by Eq. 8, where we used derived $\lambda_{e-ph}$ and bulk density of charge carriers in compressed titanium, $n$, measured by Zhang *et al* [54]. Due to Zhang *et al* [54] reported the $R(T)$ and $n$ measured at different pressures, for $T_F$ calculations we assumed the following approximations: $n(P = 18\ GPa) = n(P = 31\ GPa) = 1.72 \times 10^{28}\ m^{-3}$; $n(P = 154\ GPa) = 2.39 \times 10^{28}\ m^{-3}$; $n(P = 180\ GPa) = n(P = 183\ GPa) = n(P = 177\ GPa) = 1.70 \times 10^{28}\ m^{-3}$.

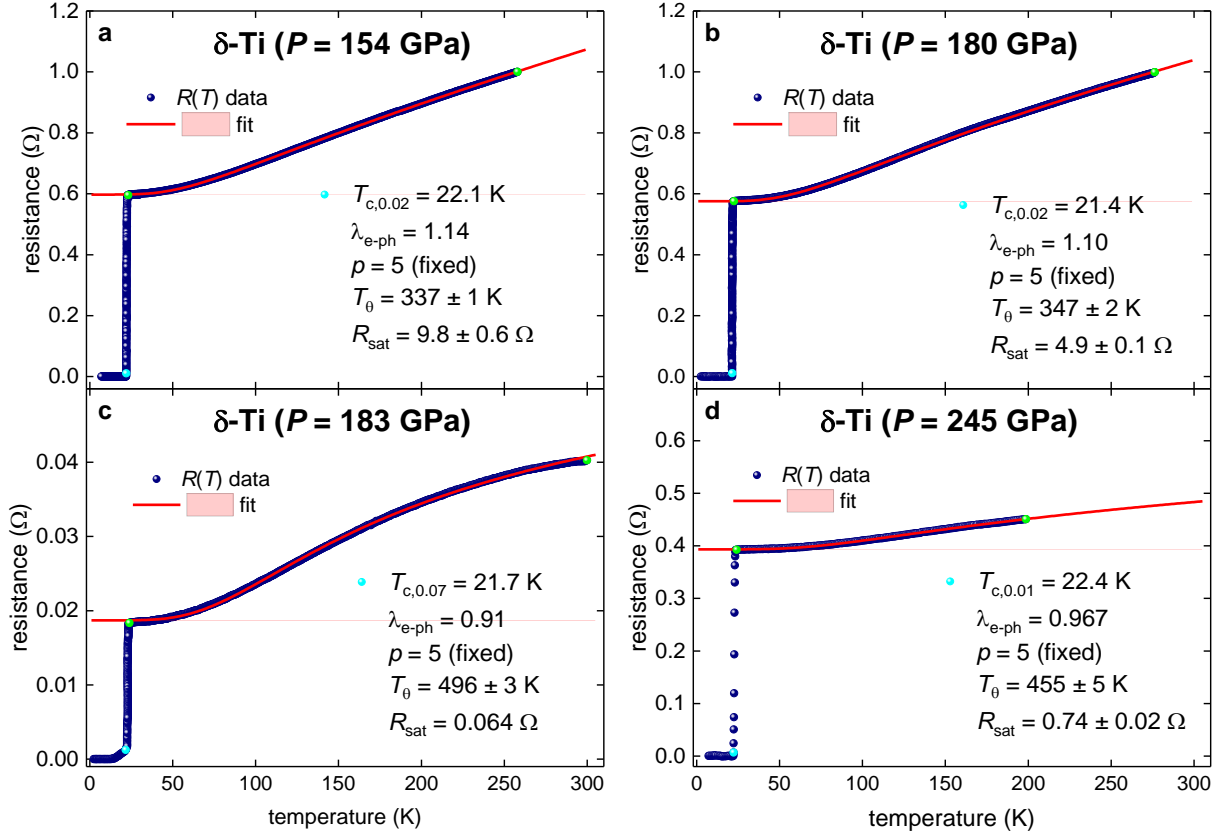

**Figure 3.** Temperature dependent resistance data, $R(T)$, for compressed titanium ($\delta - Ti$-phase) and data fit to the Debye model (Eq. 3, $p = 5$ ($fixed$)). Raw data reported by Zhang *et al* [54] (Panels a,b,d) and Liu *et al* [55] (Panel c). Green balls indicate the bounds for which $R(T)$ data was used for the fit to Eq. 3. Deduced parameters are (a) $T_\theta = 337 \pm 1\ K$, $T_{c,0.02} = 22.1\ K$, $\lambda_{e-ph} = 1.14$, fit quality is 0.99992. (b) $T_\theta = 347 \pm 2\ K$, $T_{c,0.02} = 21.4\ K$, $\lambda_{e-ph} = 1.10$, fit quality is 0.9998. (c) $T_\theta = 496 \pm 3\ K$, $T_{c,0.07} = 21.7\ K$, $\lambda_{e-ph} = 0.91$, fit quality is 0.9996. (d) $T_\theta = 455 \pm 5\ K$, $T_{c,0.01} = 22.4\ K$, $\lambda_{e-ph} = 0.967$, fit quality is 0.9997. 95% confidence bands are shown.

The evolution of the adiabaticity strength constant $\frac{T_\theta}{T_F}$ vs pressure is also showed in Figure 4,c. It can be seen (Figure 4) that there is a very good agreement between calculated by first principles calculations and extracted from experiment $\lambda_{e-ph}$ and characteristic phonon temperatures, $T_\theta$ and $\frac{1}{0.827} \times \frac{\hbar}{k_B} \times \omega_{log}$, at low and high applied pressures. More experimental data is required to perform more detailed comparison between calculated and experimental values.

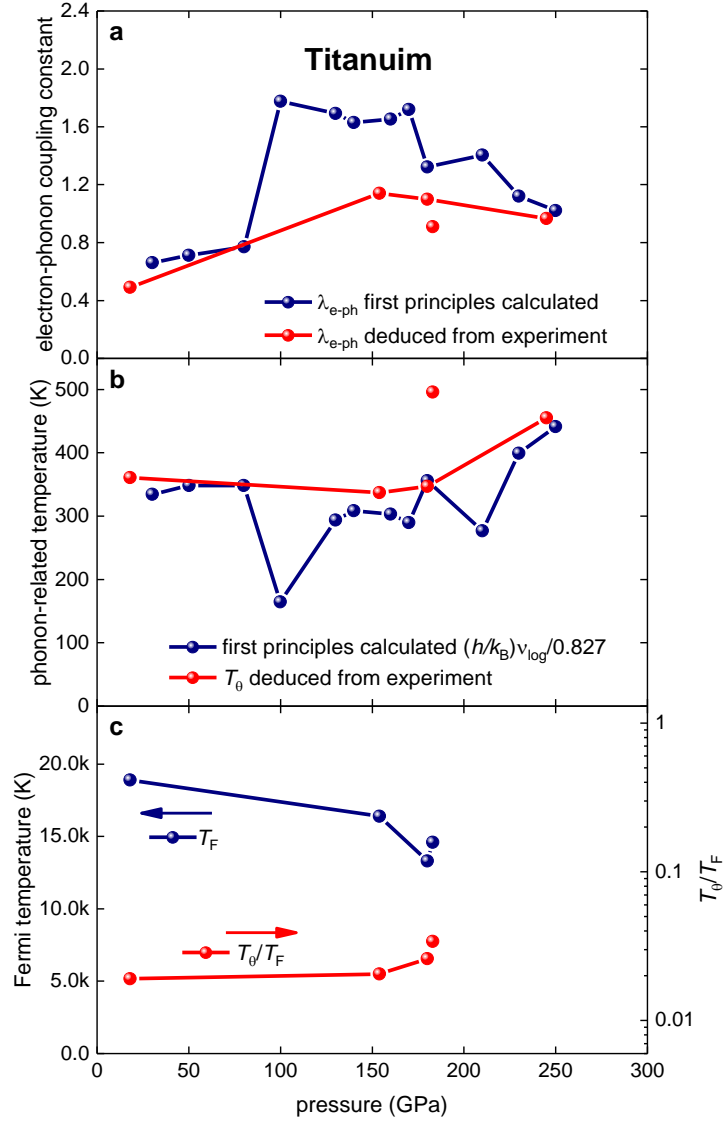

**Figure 4.** Evolution of (a) the electron-phonon coupling constant $\lambda_{e-ph}$; (b) characteristic phonon temperatures $T_\theta$ and $\frac{\hbar}{k_B}\omega_{log}$; and (c) Fermi temperature, $T_F$, calculated by Eq. 8 and the used of carrier density reported by Zhang *et al* [54] and deduced $\lambda_{e-ph}$ (in Panel (a)) and the nonadiabaticity strength constant, $\frac{T_\theta}{T_F}$, for highly compressed titanium.

Derived values for highly-compressed titanium are in Figures 5 to 7, which are widely used representation of main superconducting families (while other global scaling laws are utilized different variables [71,121–127]).

It is interesting to note, that $\delta - Ti$ is located in close proximity to A15 superconductors in all of these plots (Figures 5 to 7). It is more likely, that this is a reflection that the highest performance of the electron-phonon mediated superconductivity in metals and alloys is achieved for these materials.

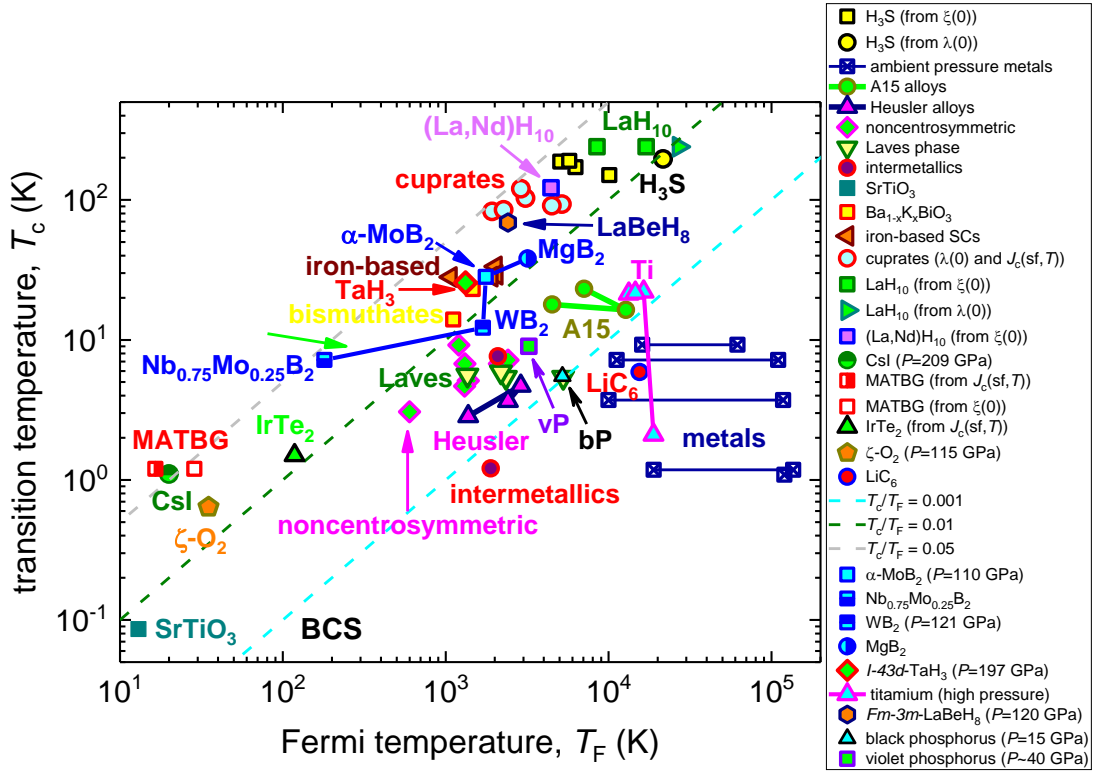

**Figure 5.** Uemura plot, where highly-compressed $Ti$, $TaH_3$, $LaBeH_8$, and black and violet phosphorous (BP and VP, respectively) are shown together with several families of superconductors: metals, iron-based superconductors, diborides, cuprates, Laves phases, hydrides, and others. References on original data can be found in Refs. [79,89,128,129].

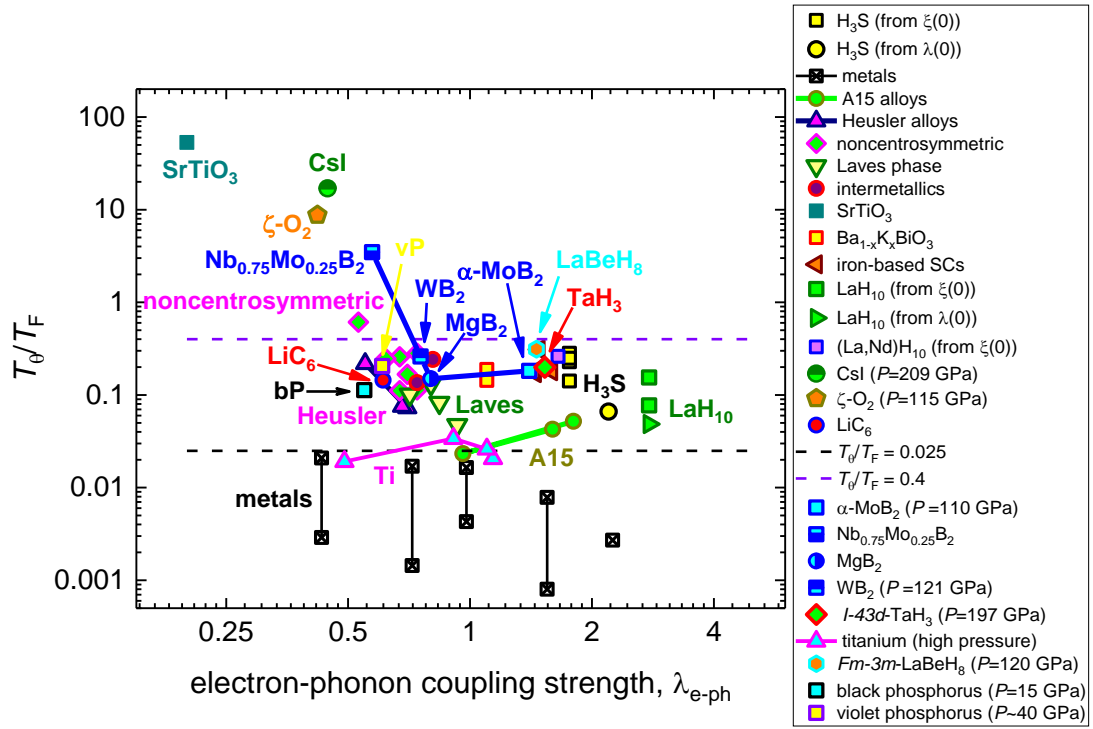

**Figure 6.** The nonadiabaticity strength constant $\frac{T_\theta}{T_F}$ vs $\lambda_{e-ph}$ where several families of superconductors and highly-compressed $Ti$, $TaH_3$, $LaBeH_8$, black and violet phosphorous are shown. References on original data can be found in Refs. [79,89,128,129].

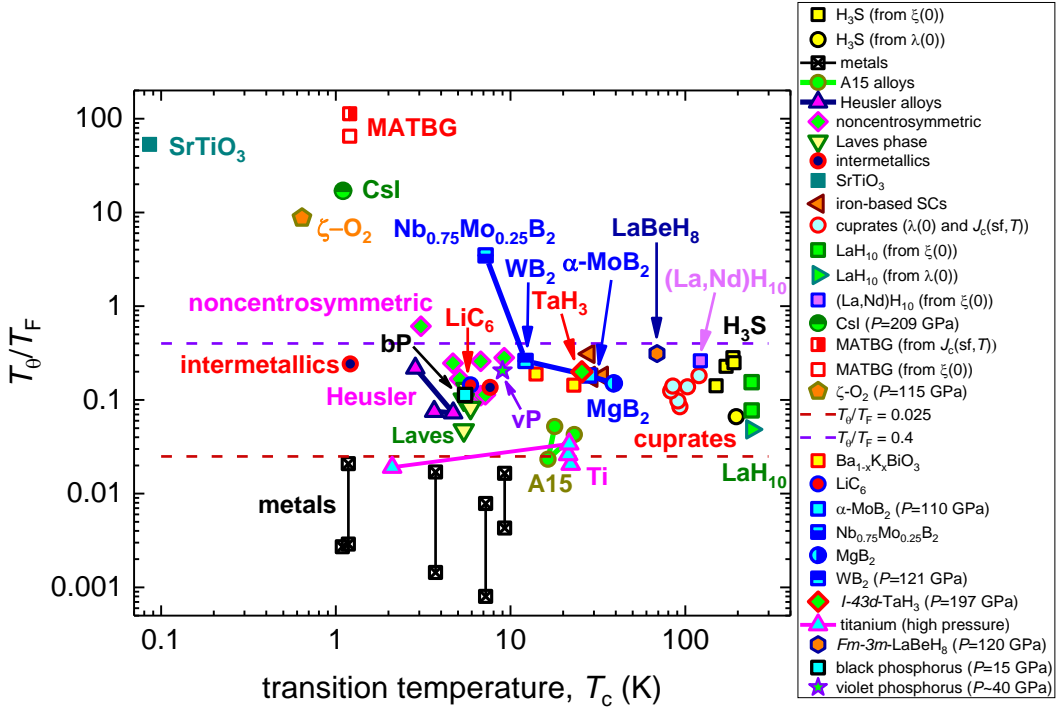

**Figure 7.** The nonadiabaticity strength constant $\frac{T_\theta}{T_F}$ vs $T_c$ for several families of superconductors and highly-compressed $Ti$, $TaH_3$, $LaBeH_8$, black and violet phosphorous are shown. References on original data can be found in Refs. [79,89,128,129].

### 3.2. Highly-compressed I-43d-phase of TaH₃

Recently, He *et al* [21] reported on the observation of high-temperature superconductivity in highly-compressed *I-43d*-phase of TaH₃. In Figure 8 we showed the fit of the $R(T)$ dataset measured by He *et al* [21] for the tantalum hydride compressed at $P = 197\ GPa$.

By utilizing Eqs. 4-6, we deduced $\lambda_{e-ph} = 1.53$ (Fig. 8), which is within ballpark value for other highly compressed hydride superconductors [3,106].

Because He *et al* [21] did not report result of Hall coefficient measurements, we deduced the Fermi temperature by the use of Eq. 9, and this we deduced $B_{c2}(T)$ dataset from $R(T,B)$ curves reported by He *et al* [21] in their Figure 2,a [21], for which we utilized the criterion of $\frac{R(T)}{R_{norm}} = 0.02$. Obtained $B_{c2}(T)$ data and data fit are shown in Figure 8(b). Deduced $\xi(0) = 5.45 \pm 0.10\ nm$.

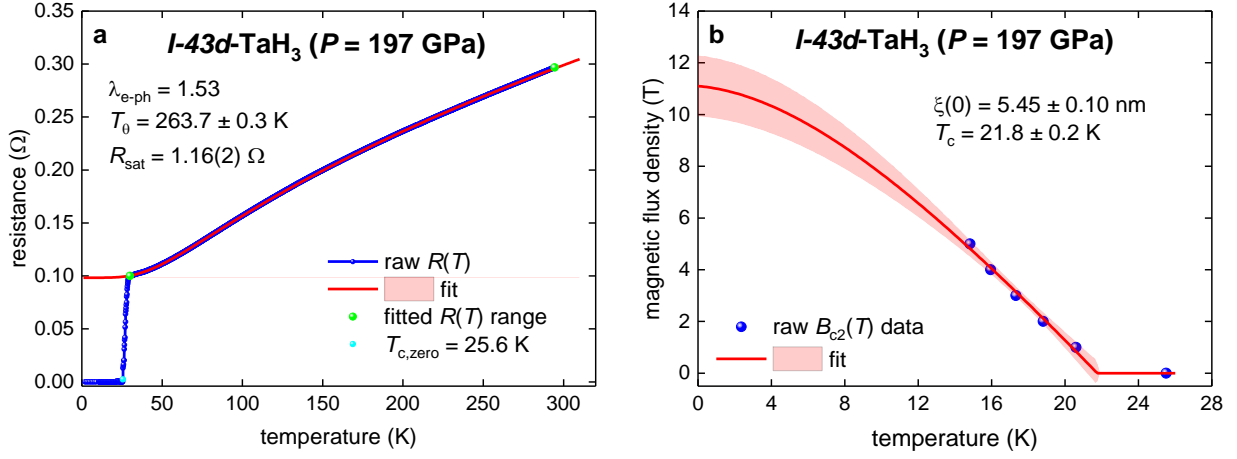

**Figure 8.** Analyzed experimental data for *I-43d*-phase of TaH₃ at $P = 197$ GPa (raw data reported by He *et al* [21]). (a) Temperature dependent resistance data, $R(T)$, and data fit to Eq. 3. Green balls indicate the bounds for which $R(T)$ data was used for the fit to Eq. 3. Deduced $T_\theta = 263.7 \pm 0.3\ K$, $T_{c,zero} = 25.6\ K$, $\lambda_{e-ph} = 1.53$, fit quality is 0.99998. (b) The upper critical field data, $B_{c2}(T)$, and data fit to Eq. 7. Definition $B_{c2}(T)$ criterion of $\frac{R(T)}{R_{norm}} = 0.02$ was used. Deduced parameters are: $\xi(0) = 2.33 \pm 0.02\ nm$, $T_c = 21.8 \pm 0.2\ K$. Fit quality is 0.9943. 95% confidence bands are shown by pink shadow areas in both panels.

To calculate the Fermi temperature in *I-43d*-phase of TaH₃ at $P = 197$ GPa, we substituted derived $\lambda_{e-ph} = 1.53$ and $\xi(0) = 5.45\ nm$ in Eq. 9, where $\alpha = \frac{2\Delta(0)}{k_B \cdot T_c} = 4.39$ was obtained by substituting $\lambda_{e-ph} = 1.53$ in Eq. 10 (Figure 1).

In the result of our analysis the following fundamental parameters of the *I-43d*-phase of TaH₃ ($P = 197\ GPa$) have been extracted:

(1) the Debye temperature, $T_\theta = 263\ K$;

(2) the electron-phonon coupling constant, $\lambda_{e-ph} = 1.53 \pm 0.13$;

(3) the ground state coherence length, $\xi(0) = 1.53 \pm 0.13$;

(4) the Fermi temperature, $T_F = 1324 \pm 74\ K$;

(5) $\frac{T_c}{T_F} = 0.019 \pm 0.01$, which implies that the this phase falls in unconventional superconductors band in the Uemura plot;

(6) the nonadiabaticity strength constant, $\frac{T_\theta}{T_F} = 0.20 \pm 0.01$.

In Figures 5-7 one can see the position of the *I-43d*-phase of TaH$_3$ at $P = 197$ GPa (within other representative materials from main families of superconductors), from which can be concluded that TaH$_3$ are typical superhydride exhibited similar strength of nonadiabatic effects to its near room temperature counterparts, i.e. H$_3$S and LaH$_{10}$.

### 3.3. Highly-compressed Fm-3m-phase of LaBeH$_8$

Recently, Song *et al* [98] reported on the observation of high-temperature superconductivity with in highly-compressed LaBeH$_8$. Crystalline structure of this superhydride at $P = 120\ GPa$ was identified as $Fm\bar{3}m$, which was predicted (as one of several possibilities) by Zhang *et al* [130]. In Figure 9(a) we showed the fit of the $R(T)$ dataset measured by Song *et al* [98] in the LaBeH$_8$ compressed at $P = 120\ GPa$.

$B_{c2}(T)$ dataset was extracted from $R(T,B)$ curves reported Song *et al* [98] in their Figure 3,a [98]. For $B_{c2}(T)$ definition we utilized the criterion of $\frac{R(T)}{R_{norm}} = 0.25$. Obtained $B_{c2}(T)$ data and data fit are shown in Figure 9(b). Deduced $\xi(0) = 2.8\ nm$.

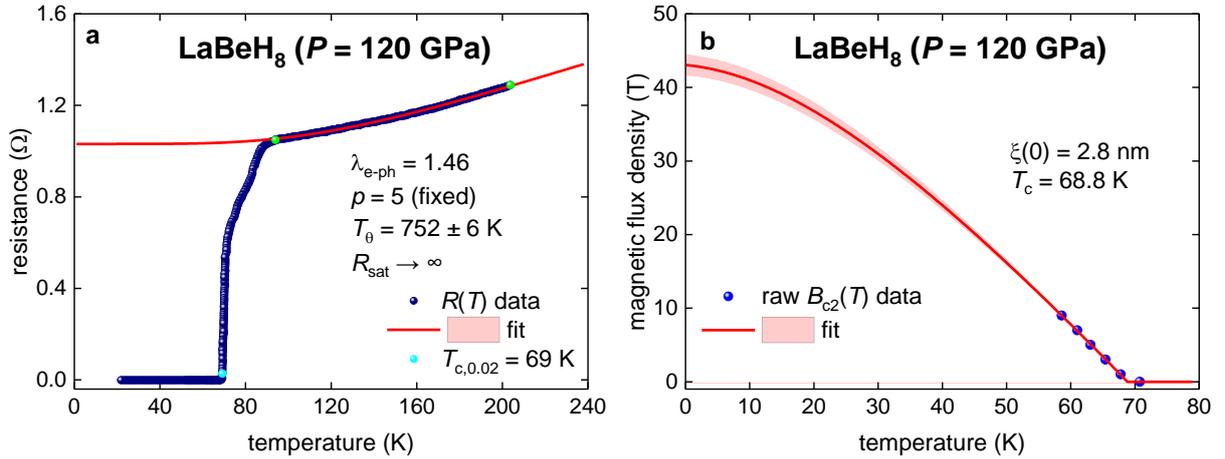

**Figure 9.** Analyzed experimental data for $Fm\bar{3}m$-phase of LaBeH$_8$ at $P = 120$ GPa (raw data reported by Song *et al* [98]). (a) Temperature dependent resistance data, $R(T)$, and data fit to Eq. 3. Green balls indicate the bounds for which $R(T)$ data was used for the fit to Eq. 3. Deduced $T_\theta = 752 \pm 6\ K$, $T_{c,0.02} = 269\ K$, $\lambda_{e-ph} = 1.46$, fit quality is 0.9990. (b) The upper critical field data, $B_{c2}(T)$, and data fit to Eq. 7. Definition $B_{c2}(T)$ criterion of $\frac{R(T)}{R_{norm}} = 0.25$ was used. Deduced parameters are: $\xi(0) = 2.8\ nm$, $T_c = 68.8\ K$. Fit quality is 0.9995. 95% confidence bands are shown by pink shadow areas in both panels.

Data analysis by the same routine described in previous Section 3.2 showed that $Fm\bar{3}m$-phase of LaBeH$_8$ at $P$ = 120 GPa exhibits the following parammeters:

(1) the Debye temperature, $T_\theta = 752 \pm 6\ K$;

(2) the electron-phonon coupling constant, $\lambda_{e-ph} = 1.46$;

(3) the ground state coherence length, $\xi(0) = 2.80 \pm 0.02\ nm$;

(4) the Fermi temperature, $T_F = 2413\ K$;

(5) $\frac{T_C}{T_F} = 0.029$, which implies that this phase falls in unconventional superconductors band in the Uemura plot;

(6) the nonadiabaticity strength constant, $\frac{T_\theta}{T_F} = 0.31 \pm 0.01$.

### 3.4. Highly-compressed black phosphorous

Impact of high pressure on the superconducting parameters of black phosphorous has studied over several decades [99–101]. Recent detailed studies in this field have reported by Guo *et al* [100] and Li *et al* [99].

To show the reliability of high-pressure studies of superconductors (which was recently questioned by non-experts in the field [131,132]) in Figure 10 we showed raw $R(T)$ datasets measured at $P = 15\ GPa$ by two independent groups, by Shirotani *et al* [101] and Li *et al* [99], whose reports have been published within a time frame of 24 years.

The agreement between deduced $\lambda_{e-ph}$ (Figure 10) from two datasets [99,101] is remarkable. It should be noted that the approach, used for this analysis (Figure 10), has been developed to analyze data measured in highly-compressed near-room temperature superconductors [106], which particularly implies that concerns expressed by non-experts in the field [131–134] in regard of highly-compressed near-room temperature hydride superconductors do not have any scientific background.

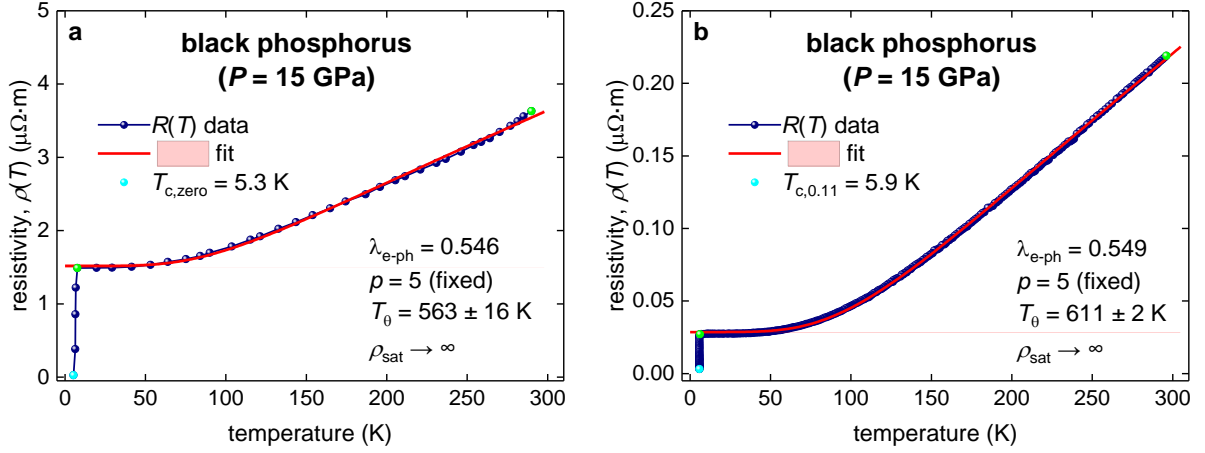

**Figure 10.** Analysis of experimental $\rho(T)$ datasets for black phosphorus compressed at $P = 15$ GPa reported by (a) Shirotani *et al* [101] and by (b) Li *et al* [99]. Green balls indicate the bounds for which $\rho(T)$ data were used for the fit to Eq. 3. Deduced parameters are: (a) $T_\theta = 563 \pm 16\ K$, $T_{c,zerp} = 5.3\ K$, $\lambda_{e-ph} = 0.546$, fit quality is 0.9983; (b) (a) $T_\theta = 611 \pm 2\ K$, $T_{c,zerp} = 5.9\ K$, $\lambda_{e-ph} = 0.549$, fit quality is 0.9998. 95% confidence bands are shown by pink shadow areas in both panels.

In Figure 11 we showed $B_{c2}(T)$ datasets extracted from raw $R(T,B)$ datasets measured at very close pressure, $P = 15.9\ GPa$ [100] and $P = 15\ GPa$ [99], which were also reported by two independent groups. For the $B_{c2}(T)$ definition we utilized the same strict criterion of $\frac{R(T)}{R_{norm}} = 0.01$ for both $R(T,B)$ datasets in Figure 11.

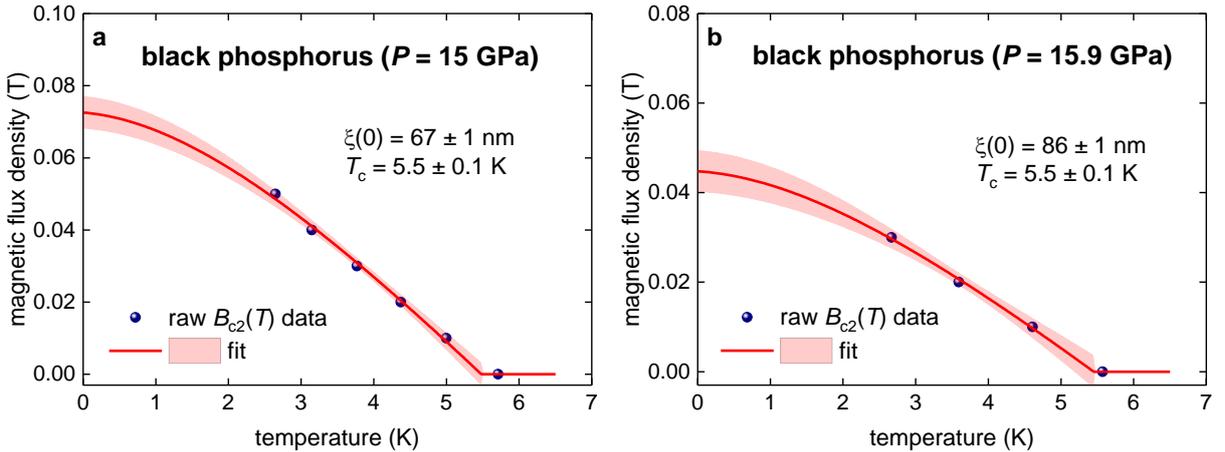

**Figure 11.** Analysis of experimental $B_{c2}(T)$ datasets for black phosphorus compressed at (a) $P = 15$ GPa reported by Li *et al* [99], and (b) $P = 15.9$ GPa reported by Guo *et al* [100]. Deduced parameters are: (a) $\xi(0) = 67 \pm 1\ K$, $T_c = 5.5 \pm 0.1\ K$, fit quality is 0.9965; (b) (a) $\xi(0) = 86 \pm 1\ K$, $T_c = 5.5 \pm 0.1\ K$, fit quality is 0.9981. 95% confidence bands are shown by pink shadow areas in both panels.

Average deduced parameters for black phosphorus $P = 15\ GPa$, which we derived from experimental data analysis reported by three different groups and which were used to position the black phosphorus in Figures 1,5-7 are:

(1) the Debye temperature, $T_\theta = 587\ K$;

(2) the electron-phonon coupling constant, $\lambda_{e-ph} = 0.548$;

(3) the ground state coherence length, $\xi(0) = 77\ nm$;

(4) the Fermi temperature, $T_F = 5200\ K$;

(5) $\frac{T_c}{T_F} = 0.001$, which implies that black phosphorus falls in conventional superconductors band in the Uemura plot;

(6) the nonadiabaticity strength constant, $\frac{T_\theta}{T_F} = 0.11$.

Deduced parameters show that the black phosphorus compressed at $P = 15\ GPa$ exhibits low strength of nonadiabatic effects.

### 3.5. Highly-compressed violet phosphorous

Recently, Wu *et al* [53] reported on the observation of the superconducting state in violet phosphorus (vP) with $T_c > 5\ K$ when the material is subjected to high pressure in the range of $3.6\ GPa \leq P \leq 40.2\ GPa$. In Figure 12(a) we showed the $R(T)$ dataset, and data fit to Eq. 3, measured by Wu *et al* [53] in the violet phosphorus compressed at $P = 40.2\ GPa$.

$B_{c2}(T)$ dataset was extracted from the only $R(T,B)$ dataset reported by Wu *et al* [53] for material compressed at Wu *et al* [53]. For $B_{c2}(T)$ definition we utilized the criterion of $\frac{R(T)}{R_{norm}} = 0.14$. Deduced $B_{c2}(T)$ dataset is shown in Figure 12(b). The fit to Equation 7 (which is single band model) has low quality, because $B_{c2}(T)$ has an upturn at $T \lesssim 4\ K$. We interpreted this upturn as an evidence for the second band opening at $T \lesssim 4\ K$, and, thus, we fitted data used two-band model [128,135]:

$$B_{c2,total}(T) = B_{c2,band1}(T) + B_{c2,band2}(T) \qquad (13)$$

where $B_{c2,band1}(T)$ and $B_{c2,band2}(T)$ exhibit their independent transition temperature and the coherence length. Deduced values are listed in the Figure Caption to Figure 12. However for further analysis we used $T_c = T_{c,band1} = 9.0\ K$ and $\xi(0)_{total} = 36 \pm 1\ nm$.

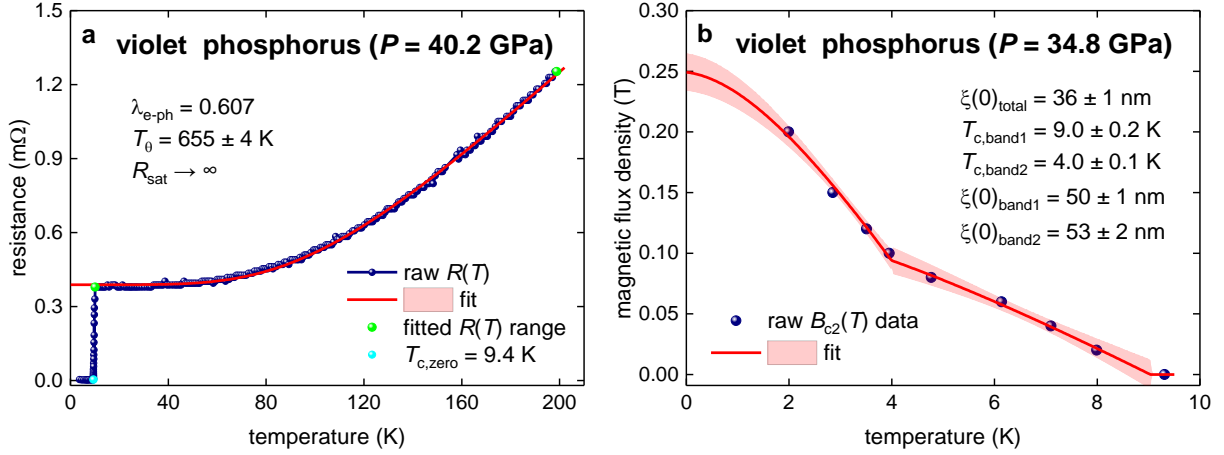

**Figure 12.** Analysis of experimental data for violet phosphorus compressed at (a) $P = 40.2\ GPa$ and (b) $P = 34.8\ GPa$. Raw data reported Wu *et al* [53]. (a) Temperature dependent resistance data, $R(T)$, and data fit to Eq. 3. Green balls indicate the bounds for which $R(T)$ data was used for the fit to Eq. 3. Deduced $T_\theta = 655 \pm 4\ K$, $T_{c,0.01} = 9.4\ K$, $\lambda_{e-ph} = 0.607$, fit quality is 0.9991. (b) The upper critical field data, $B_{c2}(T)$, and data fit to Eq. 13. Definition $B_{c2}(T)$ criterion of $\frac{R(T)}{R_{norm}} = 0.14$ was used. Deduced parameters are: $\xi(0)_{band1} = 50 \pm 1\ nm$, $T_{c,band1} = 9.0 \pm 0.2\ K$, $\xi(0)_{band1} = 53 \pm 2\ nm$, $T_{c,band2} = 4.0 \pm 0.1\ K$. Fit quality is 0.9977. 95% confidence bands are shown by pink shadow areas in both panels.

Processing data by the same approach described in previous Sections, we derived the following parameters for violet phosphorus compressed at $P \sim 40\ GPa$:

(1) the Debye temperature, $T_\theta = 665 \pm 4\ K$;

(2) the electron-phonon coupling constant, $\lambda_{e-ph} = 0.607$;

(3) the ground state coherence length, $\xi(0) = 36 \pm 1\ nm$;

(4) the Fermi temperature, $T_F = 3240\ K$;

(5) $\frac{T_c}{T_F} = 0.003$, which implies that this phase falls in close proximity to conventional superconductors band in the Uemura plot;

(6) the nonadiabaticity strength constant, $\frac{T_\theta}{T_F} = 0.21$.

Derived parameters imply that the violet phosphorus (compressed at $P \sim 40\ GPa$) exhibits moderate level of nonadiabatic effects similar to the ones in highly-compressed hydrogen-rich near-room temperature superconductors $H_3S$ and $LaH_{10}$.

## 4. Discussion

As it was mentioned above superconductors can be classified by the ratio of maximum phonon energy, $\hbar\omega_D$ (where $\omega_D$ is Debye frequency) to the charge carrier energy at the Fermi level, $\frac{\hbar\omega_D}{k_B T_F}$. For practical use, it is more convenient to replace $\hbar\omega_D$ term by $k_B T_\theta$, where $T_\theta$ is the Debye temperature, which can be deduced from experimental measurements.

Thus, in so-called adiabatic regime, $\frac{\hbar\omega_D}{k_B T_F} = \frac{T_\theta}{T_F} \lesssim 10^{-3}$, superconductors exhibit very fast charge carriers and relatively slow phonons. This condition is satisfied for pure metals and some superconducting alloys (Figures 5-7).

However, as it can be seen in Figures 6,7, more than ¾ of superconductors (including important for practical use Nb₃Sn, MgB₂, pnictides, cuprates and record high-$T_c$ near-room temperature superconducting hydrides) have the ratio in a different range [79,89]:

$$0.025 \leq \frac{\hbar\omega_D}{k_B T_F} \leq 0.4 \tag{14}$$

Our experimental data search [79,89] revealed that only six superconductors exhibit (Figures 6,6):

$$\frac{\hbar\omega_D}{k_B T_F} > 0.4 \tag{15}$$

These materials are [79,89]: Nb₀.₇₅Mo₀.₂₅B₂, Nb₀.₅Os₀.₅, highly compressed metalized oxygen, magic-angle twisted bilayer graphene, SrTiO₃, and highly compressed metalized ionic salt CsI. It should be stressed that all these superconductors exhibit low transition temperature, $T_c < 8\ K$.

In this regard, studied in this report five recently discovered superconductors (Sections 3.1-3.5) confirmed the validity of Equation 14. And, thus, perhaps a deep physical origin related to the strength of the nonadiabaticity $\frac{\hbar\omega_D}{k_B T_F} = \frac{T_\theta}{T_F}$ within a range indicated in Equation 14 can be revealed.

## 4. Conclusions

In this work, we analysed experimental data reported for five recently discovered highly-compressed superconductors: $\delta - Ti$ [54,55], $TaH_3$ [21], $LaBeH_8$ [98], black phosphorous [99–101], and violet phosphorous [53], for which we established several superconducting parameters, including the strength of nonadiabaticity, $\frac{\hbar\omega_D}{k_B T_F} = \frac{T_\theta}{T_F}$.

**Funding:** This research was funded by the Ministry of Science and Higher Education of the Russian Federation, grant number No. 122021000032-5 (theme "Pressure"). The research funding from the Ministry of Science and Higher Education of the Russian Federation (Ural Federal University Program of Development within the Priority-2030 Program) is gratefully acknowledged.

## References

1. Drozdov, A. P., Eremets, M. I., Troyan, I. A., Ksenofontov, V. & Shylin, S. I. Conventional superconductivity at 203 kelvin at high pressures in the sulfur hydride system. *Nature* **525**, 73–76 (2015).
2. Duan, D. *et al.* Pressure-induced metallization of dense (H2S)2H2 with high-Tc superconductivity. *Sci. Rep.* **4**, 6968 (2014).
3. Lilia, B. *et al.* The 2021 room-temperature superconductivity roadmap. *J. Phys. Condens. Matter* **34**, 183002 (2022).
4. Li, Y., Hao, J., Liu, H., Li, Y. & Ma, Y. The metallization and superconductivity of dense hydrogen sulfide. *J. Chem. Phys.* **140**, (2014).
5. Sun, D. *et al.* High-temperature superconductivity on the verge of a structural instability in lanthanum superhydride. *Nat. Commun.* **12**, 6863 (2021).
6. Ma, L. *et al.* High-Temperature Superconducting Phase in Clathrate Calcium Hydride $CaH_6$ up to 215 K at a Pressure of 172 GPa.

*Phys. Rev. Lett.* **128**, 167001 (2022).

7. Wang, H., Tse, J. S., Tanaka, K., Iitaka, T. & Ma, Y. Superconductive sodalite-like clathrate calcium hydride at high pressures. *Proc. Natl. Acad. Sci.* **109**, 6463–6466 (2012).

8. Bi, J. *et al.* Giant enhancement of superconducting critical temperature in substitutional alloy (La,Ce)H9. *Nat. Commun.* **13**, 5952 (2022).

9. Du, M., Song, H., Zhang, Z., Duan, D. & Cui, T. Room-Temperature Superconductivity in Yb/Lu Substituted Clathrate Hexahydrides under Moderate Pressure. *Research* **2022**, (2022).

10. Alarco, J. A., Almutairi, A. & Mackinnon, I. D. R. Progress Towards a Universal Approach for Prediction of the Superconducting Transition Temperature. *J. Supercond. Nov. Magn.* **33**, 2287–2292 (2020).

11. Kostrzewa, M., Szczęśniak, K. M., Durajski, A. P. & Szczęśniak, R. From LaH10 to room–temperature superconductors. *Sci. Rep.* **10**, 1592 (2020).

12. Drozdov, A. P. *et al.* Superconductivity at 250 K in lanthanum hydride under high pressures. *Nature* **569**, 528–531 (2019).

13. Somayazulu, M. *et al.* Evidence for Superconductivity above 260 K in Lanthanum Superhydride at Megabar Pressures. *Phys. Rev. Lett.* **122**, 027001 (2019).

14. Semenok, D. V. *et al.* Superconductivity at 161 K in thorium hydride ThH10: Synthesis and properties. *Mater. Today* **33**, 36–44 (2020).

15. Troyan, I. A. *et al.* Anomalous High-Temperature Superconductivity in YH 6. *Adv. Mater.* **33**, 2006832 (2021).

16. Semenok, D. V. *et al.* Superconductivity at 253 K in lanthanum–yttrium ternary hydrides. *Mater. Today* **48**, 18–28 (2021).

17. Kong, P. *et al.* Superconductivity up to 243 K in the yttrium-hydrogen system under high pressure. *Nat. Commun.* **12**, 5075 (2021).

18. Zhang, C. *et al.* Superconductivity in zirconium polyhydrides with Tc above 70 K. *Sci. Bull.* **67**, 907–909 (2022).

19. Zhang, C. L. *et al.* Superconductivity above 80 K in polyhydrides of hafnium. *Mater. Today Phys.* **27**, 100826 (2022).

20. Minkov, V. S., Prakapenka, V. B., Greenberg, E. & Eremets, M. I. A Boosted Critical Temperature of 166 K in Superconducting D 3 S Synthesized from Elemental Sulfur and Hydrogen. *Angew. Chemie* **132**, 19132–19136 (2020).

21. He, X. *et al.* Superconductivity Observed in Tantalum Polyhydride at High Pressure. *Chinese Phys. Lett.* **40**, 057404 (2023).

22. Minkov, V. S. *et al.* Magnetic field screening in hydrogen-rich high-temperature superconductors. *Nat. Commun.* **13**, 3194 (2022).

23. Minkov, V. S., Ksenofontov, V., Bud'ko, S. L., Talantsev, E. F. & Eremets, M. I. Magnetic flux trapping in hydrogen-rich high-temperature superconductors. *Nat. Phys.* (2023) doi:10.1038/s41567-023-02089-1.

24. Troyan, I. A. *et al.* Non-Fermi-Liquid Behavior of Superconducting SnH$_4$. *arXiv* (2023) doi:10.48550/arXiv.2303.06339.

25. Flores-Livas, J. A. *et al.* A perspective on conventional high-temperature superconductors at high pressure: Methods and materials. *Phys. Rep.* **856**, 1–78 (2020).

26. Bhattacharyya, P. *et al.* Imaging the Meissner effect and flux trapping in a hydride superconductor at megabar pressures using a nanoscale quantum sensor. (2023).

27. Goh, S. K., Zhang, W. & Yip, K. Y. Trapped magnetic flux in superconducting hydrides. *Nat. Phys.* (2023) doi:10.1038/s41567-023-02101-8.

28. Ho, K. O. *et al.* Spectroscopic Study of N- $V$ Sensors in Diamond-Based High-Pressure Devices. *Phys. Rev. Appl.* **19**, 044091 (2023).

29. Ho, K. O. *et al.* Probing the Evolution of the Electron Spin Wave Function of the Nitrogen-Vacancy Center in Diamond via Pressure Tuning. *Phys. Rev. Appl.* **18**, 064042 (2022).

30. Ho, K. O. *et al.* Probing Local Pressure Environment in Anvil Cells with Nitrogen-Vacancy (N- $V^{-}$ ) Centers in Diamond. *Phys. Rev. Appl.* **13**, 024041 (2020).

31. Yip, K. Y. *et al.* Measuring magnetic field texture in correlated electron systems under

extreme conditions. *Science (80-. ).* **366**, 1355–1359 (2019).

32. Sakata, M. *et al.* Superconductivity of lanthanum hydride synthesized using AlH 3 as a hydrogen source. *Supercond. Sci. Technol.* **33**, 114004 (2020).

33. Mozaffari, S. *et al.* Superconducting phase diagram of H3S under high magnetic fields. *Nat. Commun.* **10**, 2522 (2019).

34. Chen, W. *et al.* Synthesis of molecular metallic barium superhydride: pseudocubic BaH12. *Nat. Commun.* **12**, 1–9 (2021).

35. Zhou, D. *et al.* Superconducting praseodymium superhydrides. *Sci. Adv.* **6**, 1–9 (2020).

36. Hong, F. *et al.* Possible superconductivity at ∼70 K in tin hydride SnHx under high pressure. *Mater. Today Phys.* **22**, 100596 (2022).

37. Li, Z. *et al.* Superconductivity above 70 K observed in lutetium polyhydrides. *Sci. China Physics, Mech. Astron.* **66**, 267411 (2023).

38. Chen, W. *et al.* High-Temperature Superconducting Phases in Cerium Superhydride with a $T_c$ up to 115 K below a Pressure of 1 Megabar. *Phys. Rev. Lett.* **127**, 117001 (2021).

39. Semenok, D. V. *et al.* Effect of Magnetic Impurities on Superconductivity in LaH 10. *Adv. Mater.* **34**, 2204038 (2022).

40. Li, Z. *et al.* Superconductivity above 200 K discovered in superhydrides of calcium. *Nat. Commun.* **13**, 2863 (2022).

41. Chen, W. *et al.* Enhancement of superconducting properties in the La–Ce–H system at moderate pressures. *Nat. Commun.* **14**, 2660 (2023).

42. Purans, J. *et al.* Local electronic structure rearrangements and strong anharmonicity in YH3 under pressures up to 180 GPa. *Nat. Commun.* **12**, 1765 (2021).

43. Shao, M., Chen, W., Zhang, K., Huang, X. & Cui, T. High-pressure synthesis of superconducting clathratelike $\mathrm{Y}\mathrm{H}_4$. *Phys. Rev. B* **104**, 174509 (2021).

44. Mariappan, S. *et al.* Electronic properties of α -Mn-type non-centrosymmetric superconductor Re 5.5 Ta under hydrostatic pressure. *Supercond. Sci. Technol.* **36**, 025002 (2023).

45. Lim, J. *et al.* Creating superconductivity in WB2 through pressure-induced metastable planar defects. *Nat. Commun.* **13**, 7901 (2022).

46. Pei, C. *et al.* Pressure-induced superconductivity at 32 K in MoB2. *Natl. Sci. Rev.* **10**, (2023).

47. Li, C. *et al.* Pressure-Tuning Superconductivity in Noncentrosymmetric Topological Materials ZrRuAs. *Materials (Basel).* **15**, 7694 (2022).

48. Pei, C. *et al.* Distinct superconducting behaviors of pressurized WB2 and ReB2 with different local B layers. *Sci. China Physics, Mech. Astron.* **65**, 287412 (2022).

49. Zamyatin, D. A., Pankrushina, E. A., Streltsov, S. V. & Ponosov, Y. S. Pressure-Induced Reversible Local Structural Disorder in Superconducting AuAgTe4. *Inorganics* **11**, 99 (2023).

50. Liu, Z. Y. *et al.* Pressure-Induced Superconductivity up to 9 K in the Quasi-One-Dimensional $\mathrm{KMn}_6\mathrm{Bi}_5$. *Phys. Rev. Lett.* **128**, 187001 (2022).

51. Shimizu, K. Superconducting elements under high pressure. *Phys. C Supercond. its Appl.* **552**, 30–33 (2018).

52. Zhang, H. *et al.* Superconductivity above 12 K with possible multiband features in CsCl-type PbS. *Phys. Rev. B* **107**, 174502 (2023).

53. Wu, Y. Y. *et al.* Pressure-induced superconductivity in the van der Waals semiconductor violet phosphorus. (2023) doi:https://doi.org/10.48550/arXiv.2307.02989.

54. Zhang, C. *et al.* Record high Tc element superconductivity achieved in titanium. *Nat. Commun.* **13**, 5411 (2022).

55. Liu, X. *et al.* $T_c$ up to 23.6 K and robust superconductivity in the transition metal $\delta-$ Ti phase at megabar pressure. *Phys. Rev. B* **105**, 224511 (2022).

56. Ying, J. *et al.* Record High 36 K Transition Temperature to the Superconducting State of

Elemental Scandium at a Pressure of 260 GPa. *Phys. Rev. Lett.* **130**, 256002 (2023).

57. He, X. *et al.* Superconductivity above 30 K achieved in dense scandium. (2023) doi:https://doi.org/10.48550/arXiv.2303.01062.

58. Sun, H. *et al.* Signatures of superconductivity near 80 K in a nickelate under high pressure. *Nature* (2023) doi:10.1038/s41586-023-06408-7.

59. Errea, I. *et al.* Quantum crystal structure in the 250-kelvin superconducting lanthanum hydride. *Nature* **578**, 66–69 (2020).

60. Meninno, A. & Errea, I. Ab initio study of metastable occupation of tetrahedral sites in palladium hydrides and its impact on superconductivity. *Phys. Rev. B* **107**, 024504 (2023).

61. Meninno, A. & Errea, I. Absence of sizable superconductivity in hydrogen boride: A first-principles study. *Phys. Rev. B* **106**, 214508 (2022).

62. Belli, F., Novoa, T., Contreras-García, J. & Errea, I. Strong correlation between electronic bonding network and critical temperature in hydrogen-based superconductors. *Nat. Commun.* **12**, 5381 (2021).

63. Pickard, C. J., Errea, I. & Eremets, M. I. Superconducting Hydrides Under Pressure. *Annu. Rev. Condens. Matter Phys.* **11**, 57–76 (2020).

64. Errea, I. Superconducting hydrides on a quantum landscape. *J. Phys. Condens. Matter* **34**, 231501 (2022).

65. Hou, P., Belli, F., Bianco, R. & Errea, I. Strong anharmonic and quantum effects in $Pm\bar{3}n$ $AlH_3$ under high pressure: A first-principles stud. *Phys. Rev. B* **103**, 134305 (2021).

66. Errea, I. *et al.* High-Pressure Hydrogen Sulfide from First Principles: A Strongly Anharmonic Phonon-Mediated Superconductor. *Phys. Rev. Lett.* **114**, 157004 (2015).

67. Errea, I. *et al.* Quantum hydrogen-bond symmetrization in the superconducting hydrogen sulfide system. *Nature* **532**, 81–84 (2016).

68. Lyakhov, A. O., Oganov, A. R., Stokes, H. T. & Zhu, Q. New developments in evolutionary structure prediction algorithm USPEX. *Comput. Phys. Commun.* **184**, 1172–1182 (2013).

69. Goncharenko, I. *et al.* Pressure-Induced Hydrogen-Dominant Metallic State in Aluminum Hydride. *Phys. Rev. Lett.* **100**, 045504 (2008).

70. Talantsev, E. F. & Tallon, J. L. Universal self-field critical current for thin-film superconductors. *Nat. Commun.* **6**, 7820 (2015).

71. Talantsev, E. F., Crump, W. P. & Tallon, J. L. Universal scaling of the self-field critical current in superconductors: from sub-nanometre to millimetre size. *Sci. Rep.* **7**, 10010 (2017).

72. Park, S. *et al.* Superconductivity emerging from a stripe charge order in IrTe2 nanoflakes. *Nat. Commun.* **12**, 3157 (2021).

73. Talantsev, E. F., Crump, W. P., Storey, J. G. & Tallon, J. L. London penetration depth and thermal fluctuations in the sulphur hydride 203 K superconductor. *Ann. Phys.* **529**, 1–5 (2017).

74. Khasanov, R. *et al.* High pressure research using muons at the Paul Scherrer Institute. *High Press. Res.* **36**, 140–166 (2016).

75. Talantsev, E. F. Classifying superconductivity in compressed H3S. *Mod. Phys. Lett. B* **33**, 1950195 (2019).

76. Talantsev, E. F. Electron–phonon coupling constant and BCS ratios in LaH 10−y doped with magnetic rare-earth element. *Supercond. Sci. Technol.* **35**, 095008 (2022).

77. Khasanov, R. Perspective on muon-spin rotation/relaxation under hydrostatic pressure. *J. Appl. Phys.* **132**, (2022).

78. Grinenko, V. *et al.* Unsplit superconducting and time reversal symmetry breaking transitions in Sr2RuO4 under hydrostatic pressure and disorder. *Nat. Commun.* **12**, 3920 (2021).

79. Talantsev, E. F. D-Wave Superconducting Gap Symmetry as a Model for Nb1−xMoxB2 (x = 0.25; 1.0) and WB2 Diborides. *Symmetry (Basel).* **15**, 812 (2023).

80. Migdal, A. B. Interaction between Electrons and Lattice Vibrations in a Normal Metal. *Sov. Phys.–JETP* **7**, 996 (1958).

81. G. M. Eliashberg. Interactions between Electrons and Lattice Vibrations in a Superconductor. *Sov. Phys.–JETP* **11**, 696 (1960).

82. Gor'kov, L. P. Phonon mechanism in the most dilute superconductor n -type SrTiO 3. *Proc. Natl. Acad. Sci.* **113**, 4646–4651 (2016).

83. Takada, Y. Plasmon Mechanism of Superconductivity in Two- and Three-Dimensional Electron Systems. *J. Phys. Soc. Japan* **45**, 786–794 (1978).

84. Lin, X., Zhu, Z., Fauqué, B. & Behnia, K. Fermi Surface of the Most Dilute Superconductor. *Phys. Rev. X* **3**, 021002 (2013).

85. Szczęśniak, D. & Drzazga-Szczęśniak, E. A. Non-adiabatic superconductivity in the electron-doped graphene. *EPL (Europhysics Lett.* **135**, 67002 (2021).

86. Drzazga-Szczęśniak, E. A., Szczęśniak, D., Kaczmarek, A. Z. & Szczęśniak, R. Breakdown of Adiabatic Superconductivity in Ca-Doped h-BN Monolayer. *Condens. Matter* **7**, 60 (2022).

87. Szczęśniak, D., Kaczmarek, A. Z., Drzazga-Szczęśniak, E. A. & Szczęśniak, R. Phonon-mediated superconductivity in bismuthates by nonadiabatic pairing. *Phys. Rev. B* **104**, 094501 (2021).

88. Szcze śniak, D. Scalability of non-adiabatic effects in lithium-decorated graphene superconductor. *Europhys. Lett.* **142**, 36002 (2023).

89. Talantsev, E. F. Quantifying Nonadiabaticity in Major Families of Superconductors. *Nanomaterials* **13**, 71 (2022).

90. Yoon, H. *et al.* Low-density superconductivity in SrTiO$_3$ bounded by the adiabatic criterion. (2021) doi:https://doi.org/10.48550/arXiv.2106.10802.

91. Pietronero, L., Strässler, S. & Grimaldi, C. Nonadiabatic superconductivity. I. Vertex corrections for the electron-phonon interactions. *Phys. Rev. B* **52**, 10516–10529 (1995).

92. Grimaldi, C., Pietronero, L. & Strässler, S. Nonadiabatic superconductivity. II. Generalized Eliashberg equations beyond Migdal's theorem. *Phys. Rev. B* **52**, 10530–10546 (1995).

93. Cappelluti, E., Ciuchi, S., Grimaldi, C., Pietronero, L. & Strässler, S. High $T_c$ Superconductivity in $MgB_2$. *Phys. Rev. Lett.* **88**, 117003 (2002).

94. Grimaldi, C., Cappelluti, E. & Pietronero, L. Isotope effect on m * in high- T c materials due to the breakdown of Migdal's theorem. *Europhys. Lett.* **42**, 667–672 (1998).

95. Pietronero, L., Boeri, L., Cappelluti, E. & Ortenzi, L. Conventional/unconventional superconductivity in high-pressure hydrides and beyond: insights from theory and perspectives. *Quantum Stud. Math. Found.* **5**, 5–21 (2018).

96. Eremets, M. I., Shimizu, K., Kobayashi, T. C. & Amaya, K. Metallic CsI at Pressures of up to 220 Gigapascals. *Science (80-. ).* **281**, 1333–1335 (1998).

97. Talantsev, E. F. Fermi-Liquid Nonadiabatic Highly Compressed Cesium Iodide Superconductor. *Condens. Matter* **7**, 65 (2022).

98. Song, Y. *et al.* Stoichiometric Ternary Superhydride $LaBeH_8$ as a New Template for High-Temperature Superconductivity at 110 K under 80 GPa. *Phys. Rev. Lett.* **130**, 266001 (2023).

99. Li, X. *et al.* Pressure-induced phase transitions and superconductivity in a black phosphorus single crystal. *Proc. Natl. Acad. Sci.* **115**, 9935–9940 (2018).

100. Guo, J. *et al.* Electron-hole balance and the anomalous pressure-dependent superconductivity in black phosphorus. *Phys. Rev. B* **96**, 224513 (2017).

101. Shirotani, I. *et al.* Phase transitions and superconductivity of black phosphorus and phosphorus-arsenic alloys at low temperatures and high pressures. *Phys. Rev. B* **50**, 16274–16278 (1994).

102. Bardeen, J., Cooper, L. N. & Schrieffer, J. R. Theory of Superconductivity. *Phys. Rev.* **108**, 1175–1204 (1957).

103. McMillan, W. L. Transition Temperature of Strong-Coupled Superconductors. *Phys. Rev.* **167**, 331–344 (1968).

104. Dynes, R. C. McMillan's equation and the Tc of superconductors. *Solid State Commun.* **10**, 615–618 (1972).

105. Allen, P. B. & Dynes, R. C. Transition temperature of strong-coupled superconductors

reanalyzed. *Phys. Rev. B* **12**, 905–922 (1975).

106. Talantsev, E. F. Advanced McMillan's equation and its application for the analysis of highly-compressed superconductors. *Supercond. Sci. Technol.* **33**, 094009 (2020).

107. Bloch, F. Zum elektrischen Widerstandsgesetz bei tiefen Temperaturen. *Zeitschrift f�r Phys.* **59**, 208–214 (1930).

108. Grüneisen, E. Die Abhängigkeit des elektrischen Widerstandes reiner Metalle von der Temperatur. *Ann. Phys.* **408**, 530–540 (1933).

109. Fisk, Z. & Webb, G. W. Saturation of the High-Temperature Normal-State Electrical Resistivity of Superconductors. *Phys. Rev. Lett.* **36**, 1084–1086 (1976).

110. Wiesmann, H. *et al.* Simple Model for Characterizing the Electrical Resistivity in $A-15$ Superconductors. *Phys. Rev. Lett.* **38**, 782–785 (1977).

111. Carbotte, J. P. Properties of boson-exchange superconductors. *Rev. Mod. Phys.* **62**, 1027–1157 (1990).

112. Helfand, E. & Werthamer, N. R. Temperature and Purity Dependence of the Superconducting Critical Field, $H_{c2}$ . II. *Phys. Rev.* **147**, 288–294 (1966).

113. Werthamer, N. R., Helfand, E. & Hohenberg, P. C. Temperature and Purity Dependence of the Superconducting Critical Field, $H_{c2}$ . III. Electron Spin and Spin-Orbit Effects. *Phys. Rev.* **147**, 295–302 (1966).

114. Baumgartner, T. *et al.* Effects of neutron irradiation on pinning force scaling in state-of-the-art Nb 3 Sn wires. *Supercond. Sci. Technol.* **27**, 015005 (2014).

115. P. Poole, C. Properties of the normal metallic state. in *Handbook of Superconductivity* 29–41 (Elsevier, 2000). doi:10.1016/B978-012561460-3/50003-5.

116. Poole, C. P., Farach, H., Creswick, R. & Prozorov, R. *Superconductivity*. (Academic Press, 2007).

117. Talantsev, E. F. The Compliance of the Upper Critical Field in Magic-Angle Multilayer Graphene with the Pauli Limit. *Materials (Basel).* **16**, 256 (2022).

118. Ho, C. Y., Powell, R. W. & Liley, P. E. Thermal Conductivity of the Elements. *J. Phys. Chem. Ref. Data* **1**, 279–421 (1972).

119. Talantsev, E. F. The dominance of non-electron–phonon charge carrier interaction in highly-compressed superhydrides. *Supercond. Sci. Technol.* **34**, 115001 (2021).

120. Semenok, D. Computational design of new superconducting materials and their targeted experimental synthesis. *PhD Thesis; Sk. Inst. Sci. Technol.* (2022) doi:10.13140/RG.2.2.28212.12161.

121. Harshman, D. R. & Fiory, A. T. High-TC Superconductivity in Hydrogen Clathrates Mediated by Coulomb Interactions Between Hydrogen and Central-Atom Electrons. *J. Supercond. Nov. Magn.* **33**, 2945–2961 (2020).

122. Homes, C. C. *et al.* A universal scaling relation in high-temperature superconductors. *Nature* **430**, 539–541 (2004).

123. Koblischka, M. R. & Koblischka-Veneva, A. Calculation of Tc of Superconducting Elements with the Roeser–Huber Formalism. *Metals (Basel).* **12**, 337 (2022).

124. Dew-Hughes, D. Flux pinning mechanisms in type II superconductors. *Philos. Mag.* **30**, 293–305 (1974).

125. Kramer, E. J. Scaling laws for flux pinning in hard superconductors. *J. Appl. Phys.* **44**, 1360–1370 (1973).

126. Talantsev, E. F. New Scaling Laws for Pinning Force Density in Superconductors. *Condens. Matter* **7**, 74 (2022).

127. Godeke, A., Haken, B. ten, Kate, H. H. J. ten & Larbalestier, D. C. A general scaling relation for the critical current density in Nb3Sn. *Supercond. Sci. Technol.* **19**, R100–R116 (2006).

128. Talantsev, E. F., Mataira, R. C. & Crump, W. P. Classifying superconductivity in Moiré graphene superlattices. *Sci. Rep.* **10**, 212 (2020).

129. Talantsev, E. F. Classifying hydrogen-rich superconductors. *Mater. Res. Express* **6**, 106002

(2019).

130.    Zhang, Z. *et al.* Design Principles for High-Temperature Superconductors with a Hydrogen-Based Alloy Backbone at Moderate Pressure. *Phys. Rev. Lett.* **128**, 047001 (2022).

131.    Hirsch, J. E. & Marsiglio, F. On Magnetic Field Screening and Expulsion in Hydride Superconductors. *J. Supercond. Nov. Magn.* **36**, 1257–1261 (2023).

132.    Hirsch, J. E. & Marsiglio, F. Evidence Against Superconductivity in Flux Trapping Experiments on Hydrides Under High Pressure. *J. Supercond. Nov. Magn.* **35**, 3141–3145 (2022).

133.    Hirsch, J. E. Enormous Variation in Homogeneity and Other Anomalous Features of Room Temperature Superconductor Samples: A Comment on Nature 615, 244 (2023). *J. Supercond. Nov. Magn.* (2023) doi:10.1007/s10948-023-06593-6.

134.    Hirsch, J. E. Electrical Resistance of Hydrides Under High Pressure: Evidence of Superconductivity or Confirmation Bias? *J. Supercond. Nov. Magn.* (2023) doi:10.1007/s10948-023-06594-5.

135.    Talantsev. Classifying Induced Superconductivity in Atomically Thin Dirac-Cone Materials. *Condens. Matter* **4**, 83 (2019).